\documentstyle[aps,multicol,graphicx,psfrag]{revtex}
\begin{document}
\draft
\widetext
\title{Thermodynamics and optical conductivity of unconventional spin density
waves}
\author{Bal\'azs D\'ora}
\address{Department of Physics, Technical University of Budapest, H-1521
Budapest, Hungary}
\author{Attila Virosztek}
\address{Department of Physics, Technical University of Budapest, H-1521
Budapest, Hungary\\
and Research Institute for Solid State Physics and Optics, P.O.Box 49,
H-1525 Budapest, Hungary}
\date{\today}
\maketitle
\widetext
\begin{abstract}
We consider the possibility of formation of an unconventional spin density
wave (USDW) in quasi-one dimensional electronic systems. In analogy with
unconventional superconductivity, we develop a mean field theory of SDW
allowing for the momentum dependent gap $\Delta({\bf k})$ on the Fermi
surface. Conditions for the appearence of such a low temperature phase are
investigated.
The excitation spectrum and basic thermodynamic properties of the
model are found to be very similar to those of $d$-wave superconductors
in spite of the different topology of their Fermi surfaces. Several
correlation functions are calculated, and
the frequency dependent conductivity
is evaluated for various gap
functions. The latter is found to reflect the maximum gap value, however
with no sharp onset for absorbtion.
\end{abstract}
%\begin{multicols}{2}      %    for
%\narrowtext               % twocolumn

\section{Introduction}

As a result of intense research during the past few decades, much
is known about the properties of density waves possessing a constant
$\Delta$ order parameter \cite{Gruner}. On the other hand it is well
known that
wavevector dependent order parameters $\Delta({\bf k})$ taking different
values at different points on the Fermi surface play an important role in
theories of superconductivity (termed unconventional)\cite{AM}
in various systems including the high-$T_c$
cuprates \cite{Maki}. Therefore the need to work out the theory of
unconventional
density waves follows naturally from these precedents.

This project however
is not just of academic interest. The anomalously small magnetic moment in
URu$_2$Si$_2$\cite{Isaacs} was suggested to be explained among other
possibilities\cite{BarGor,Shah} by unconventional SDW on a
quasi-two dimensional (square) Fermi surface \cite{IO}. The basis of
this suggestion is the fact that an USDW is in fact not a spin density
wave at all, since it is not accompanied by periodic spin density
modulation. The order parameter of the phase transition is not the
spin density itself, but another well defined quantity related to the spin
density much the same way as the "effective density" responsible for
electronic Raman scattering is related to the density
operator\cite{AbrGen,Multiband}.
As another signature of an unconventional density wave (DW),
low energy
excitations due to a DW gap vanishing on certain subsets of
the Fermi surface may be responsible for the absence of a clear optical
gap in the reflectivity data of some of the heavy-electron
materials\cite{Degiorgi}, including
URu$_2$Si$_2$. Optical data with similar features
in the low temperature phase of quasi-one dimensional Mo$_4$O$_{11}$
indicate again the possibility of an unconventional charge density wave (UCDW)
state\cite{MoO1,MoO2}.

Other candidates for unconventional DW
states include the organic conductor $\alpha$-(BEDT-TTF)$_2$KHg(SCN)$_4$, where
in spite of a clear indication for a phase transition in magnetotransport
measurements neither charge nor magnetic order has been
established\cite{Christ,Proust,Kartsovnik},
the transition-metal dichalcogenide
2$H$-TaSe$_2$ with its momentum dependent CDW gap inferred from angle resolved
photoemission studies\cite{TaSe2}, and the tetrachalcogenide (TaSe$_4$)$_2$I,
for which the magnetic susceptibility 
above the conventional CDW transition temperature shows pseudogap
behavior\cite{Johnston,Kriza}
without any observable long range charge order. Indeed, a recent
suggestion to understand the pseudogap phase of the cuprate superconductors
also
invokes the existence of an unconventional density wave (UDW)
of $d$-symmetry\cite{Sudip}.
Motivated partly by the rich phase diagram of the high $T_c$ cuprates,
significant steps have already been made towards the theory of UDW
in a two dimensional electron system\cite{Bouis}, typically on
a square lattice\cite{Nayak}.

The objective of the present paper is to develop a detailed theory of
UDW-s in quasi-one dimensional interacting electron
systems. Clearly, the topology of the Fermi surface is radically different
in this problem than in previous treatments, moreover the strong anisotropy
of transport properties in, and perpendicular to the linear chain direction
is of particular interest. A preliminary report of some of our results has
already been published\cite{Ecrys}. The article is organized as follows:
in Sec. II. we define our model, develop its thermodynamics in mean field
theory, and determine conditions for the appearance of USDW. Sec. III. is
devoted to the calculation of the most important correlation functions of
the system, while in Sec. IV. we pay particular attention to the optical
conductivity of the model. Our conclusions are given in Sec. V.

\section{Thermodynamics of the model}

We start with a quasi-one dimensional
interacting electron system described by the following one band
Hamiltonian:
\begin{equation}
 H=\sum_{\bf k,\sigma}\varepsilon({\bf k})a_{\bf k,\sigma}^{+}a_{\bf
 k,\sigma}+ \frac{1}{2V}
   \sum_{\begin{array}{c}
          {\bf k,k^\prime,q} \\
          \sigma,\sigma^\prime
         \end{array}} \tilde{V}({\bf k,k^\prime,q})a_{\bf k+q,\sigma}^{+}
         a_{\bf k,\sigma}a_{\bf k^\prime-q,\sigma^\prime}^{+}a_{\bf
         k^\prime,\sigma^\prime} ,
\label{hamilton}
\end{equation}
where $a_{\bf k,\sigma}^{+}$ and $a_{\bf k,\sigma}$ are 
creation and annihilation operators of an electron of momentum $\bf k$ and 
spin $\sigma$, $V$ is the volume of the sample and
the kinetic energy spectrum on an orthorombic lattice
\begin{equation}
\epsilon({\bf k})=-2t_{a}\cos(k_{x}a)-2t_{b}\cos(k_{y}b)-2t_{c}\cos(k_{z}c)
-\mu\label{disp}
\end{equation}
is highly anisotropic ($t_a\gg t_b, t_c$).
The interaction potential matrix elements in (\ref{hamilton}) are: 
\begin{eqnarray}
 \frac{1}{V}\tilde{V}({\bf k,k^\prime,q})=\int d^{3}r
\int d^{3}r^\prime\bar{\varphi}_{\bf
  k+q}({\bf r}) \bar{\varphi}_{\bf k^\prime-q}({\bf r^\prime}) V({\bf r-r^\prime
})
  \varphi_{\bf k^{'}}({\bf r^\prime}) \varphi_{\bf k}({\bf r}),\label{int}
\end{eqnarray}
where $\varphi_{\bf k}$ is Bloch function.
In Wannier basis $\varphi_{\bf k}(\bf r)=\it \frac{1}{\sqrt{N}}\sum_{\bf
 R} e^{\bf ikR} \varphi({\bf r-R})$, where $N$ is the number of cells and
$\varphi(\bf r)$ is the
corresponding
Wannier function
assumed to be real and
an eigenfunction of parity. We note here that in a tight-binding
solid the Wannier function is well localized, leading to a significant
dependence of the interaction matrix element (\ref{int}) on the incoming
electron momenta ${\bf k}$ and ${\bf k^\prime}$. This turns out to be
crucial in order to form an UDW, and is readily seen from the expansion
including on site and nearest neighbor two center integrals:
\begin{eqnarray}
\frac{N}{V}\tilde{V}({\bf k,k^\prime,q,\sigma,\sigma^\prime})&&=
\delta_{-\sigma,\sigma^\prime}\{U+\sum_{i}[2V_{i}\cos
q_{i}\delta_{i}+2J_{i}\cos(k_{i}-k_{i}^\prime+q_{i})\delta_{i}
+2F_{i}\cos(k_{i}^{'}+k_{i})\delta_{i}+\nonumber\\
&&+2C_{i}(\cos k_{i}\delta_{i}+\cos k_{i}^\prime\delta_{i}+
\cos(k_{i}^\prime-q_{i})\delta_{i}+
\cos(k_{i}+q_{i})\delta_{i})]\}+ \nonumber \\
&&+\delta_{\sigma,\sigma^\prime}\sum_{i}(V_{i}-J_{i})(\cos q_{i}\delta_{i}-
 \cos(k_{i}-k_{i}^\prime+q_{i})\delta_{i}).\label{spinpot}
\end{eqnarray}
In the above formula the antisymmetrized (therefore spin dependent) interaction
is given with $i=x,y,z$ and $\delta_{i}=a,b,c$. The (at most) two center
integrals are the on site Hubbard repulsion and the nearest neighbor direct,
exchange, pair-hopping and bond-charge terms
\begin{mathletters}
\label{allHubints}
\begin{eqnarray}
U=&&\int d^{3}r \int d^{3}r^\prime|\varphi({\bf r})|^{2} V({\bf r-r^\prime})
|\varphi({\bf r^\prime})|^{2},\label{Hubinta}\\
V_{i}=&&\int d^{3}r\int d^{3}r^\prime|\varphi({\bf r})|^{2} V({\bf
r-r^\prime})|\varphi({\bf r^\prime-e_{\it i}})|^{2},\label{Hubintb}\\
J_{i}=&&\int d^{3}r\int d^{3}r^\prime\bar{\varphi}({\bf
r})\bar{\varphi}({\bf r^\prime-e_{\it i}}) V({\bf r-r^\prime}) \varphi({\bf
r^\prime}) \varphi({\bf r-e_{\it i}}),\label{Hubintc}\\
F_{i}=&&\int d^{3}r \int d^{3}r^\prime\bar{\varphi}(\bf r)\it
\bar{\varphi}(\bf r^\prime)
   \it V(\bf r-r^\prime)\it \varphi(\bf r^\prime-e_{\it i})\it \varphi(\bf
   r-e_{\it i}),\label{Hubintd}\\
C_{i}=&&\int d^{3}r\int d^{3}r^\prime\bar{\varphi}(\bf r)\it
\bar{\varphi}(\bf r^\prime)
 \it V(\bf r-r^\prime)\it \varphi(\bf r^\prime)\it \varphi(\bf r-e_{\it i}),
\label{Hubinte}
\end{eqnarray}
\end{mathletters}
where ${\bf e}_i$ is the lattice vector in the $i$ direction.

Although due to its rich structure the interaction in Eq.(\ref{spinpot}) is
able to support a variety of low temperature phases\cite{Ozaki}
depending on the
Hubbard integrals in Eq.(\ref{allHubints}), we are now interested in
constructing the mean field theory of an USDW. The best nesting vector for
the spectrum (\ref{disp}) is obviously ${\bf Q}=(2k_{F},\pi/b,\pi/c)$
with the Fermi wavenumber $k_{F}$ satisfying
$\mu=-2t_a\cos(ak_F)$. In a DW expectation values of the type
$<a_{\bf k,\sigma}^{+}a_{\bf k+Q,\sigma}>$ will not vanish, defining the
order parameter of the low temperature phase
$\Delta({\bf k},\sigma)=|\Delta({\bf k},\sigma)|e^{i\phi(\bf k,\sigma)}$
as
\begin{equation}
\Delta(\bf k,\sigma)=\it \frac{1}{V}\sum_{{\bf k^\prime},\sigma^\prime}^{\prime
\prime}
\overline{\tilde{V}(\bf k^\prime,k,Q,\sigma,\sigma^\prime)}\it <a_{\bf k^\prime,
\sigma^\prime}^{+}a_{\bf
k^\prime+Q,\sigma^\prime}>.
\label{ordpar}
\end{equation}
Here the overline indicates complex conjugation and the double prime over the
summation sign restricts $k_x$ values to the interval from $-2k_F$ to $0$.
Then the mean field Hamiltonian is diagonalized in the usual way, giving
rise to a two band quasiparticle spectrum over the new Brillouin zone
($0<k_x<2k_F$) given by
\begin{equation}
 E_{\pm}({\bf k})=\frac{\varepsilon({\bf k})+\varepsilon({\bf k-Q})}{2} \pm
 \sqrt{\left(\frac{\varepsilon({\bf k})-\varepsilon({\bf
 k-Q})}{2}\right)^{2}+|\Delta({\bf k},\sigma)|^{2}}.\label{Espectrum}
\end{equation}
The new (effectively noninteracting) fermionic quasiparticles are expressed
by the original electrons as
\begin{eqnarray}
d_{+ \bf k \sigma}=e^{-i\phi(\bf k,\sigma)} u({\bf k,\sigma})
a_{\bf k,\sigma}+ v({\bf k,\sigma}) a_{\bf k-Q,\sigma},\nonumber \\
d_{- \bf k \sigma}=e^{-i\phi(\bf k,\sigma)} v({\bf k,\sigma})
a_{\bf k,\sigma}- u({\bf k,\sigma}) a_{\bf k-Q,\sigma},\label{d-a}
\end{eqnarray}
with
\begin{equation}
 \begin{array}{c}
    u({\bf k,\sigma}) \\ v({\bf k,\sigma})
 \end{array}
 = \sqrt{\frac{1}{2}\left(1 \pm \frac{\frac{\varepsilon({\bf k})-
\varepsilon({\bf k-Q})}{2}}
 {\sqrt{(\frac{\varepsilon({\bf k})-\varepsilon({\bf k-Q})}{2})^{2}+
|\Delta({\bf k,\sigma})|^{2}}}\right)}.\label{uv}
\end{equation}
Eqs.(\ref{ordpar}) and (\ref{d-a}) lead to a self consistency condition for
the order parameter known as the gap equation
\begin{equation}
 \Delta({\bf l,\sigma^\prime})=\sum_{\bf k,\sigma}^\prime\frac{1}{V}
 \overline{\tilde{V}({\bf
 k-Q,l,Q,\sigma^\prime,\sigma})}\frac{\Delta({\bf
 k,\sigma})\{f[E_{+}({\bf k,\sigma})]-f[E_{-}({\bf k,\sigma})]\}
}{2\sqrt{(\frac{\varepsilon({\bf k})-\varepsilon({\bf
 k-Q})}{2})^{2}+|\Delta({\bf k,\sigma})|^{2}}},\label{gapeq1}
\end{equation}
where the prime indicates that the ${\bf k}$ sum runs over the new Brillouin
zone only, and $f(E)$ is the Fermi function.

In the followings we suppress the spin index of the order parameter, since
in order to describe an SDW with polarization vector parallel to the $z$
axis of the spin space, we can utilize the relation
$\Delta({\bf k},+)=-\Delta({\bf k},-)=\Delta({\bf k})$. Moreover,
the structure of the gap equation makes it clear that the relevant
wavenumber values in the arguments of both the gap and the interaction are
confined to a narrow region near the Fermi sheet at $+k_F$. Therefore the
gap equation is in fact an integral equation on the $k_x=k_F$ plane of
variables $k_y$ and $k_z$ only. Making use of the electron spectrum
(\ref{disp}) linearized in $k_x$ around $k_F$:
$\xi({\bf k})=v_{F}(k_{x}-k_{F})-2t_{b}\cos(k_{y}b)-2t_{c}\cos(k_{z}c)$,
we obtain a simplified gap equation ($v_F=2at_a\sin(ak_F)$ is the Fermi
velocity):
\begin{equation}
\Delta({\bf l})=\sum_{\bf k}^\prime\frac{1}{V}\overline{P({\bf
 k,l})}\frac{\Delta({\bf k})\tanh\{{\beta\over 2}
E({\bf k})\}
 }{2E({\bf k})}
\approx\int_{-\pi/b}^{\pi/b}{dk_y\over 2\pi}\int_{-\pi/c}^{\pi/c}
{dk_z\over 2\pi}\int_0^{v_Fk_F}{d\xi\over 2\pi v_F}\overline{P({\bf k,l})}
\Delta({\bf k}){\tanh\{{\beta\over 2}\sqrt{\xi^2+|\Delta({\bf k})|^{2}}\}
\over\sqrt{\xi^2+|\Delta({\bf k})|^{2}}}.\label{gapeq}
\end{equation}
Here $E({\bf k})=\sqrt{[\xi({\bf k})]^2+|\Delta({\bf k})|^{2}}$,
$\beta=1/k_BT$, we have neglected terms of order $(t_{b,c}/t_a)^2$ in the
second expression, and the kernel of the integral equation is
\begin{equation}
\frac{P({\bf k,l})}{V}=\frac{P_{0}}{N}+\frac{P_{1}}{N}\cos(k_{y}b)
\cos(l_{y}b)+
\frac{P_{2}}{N}\sin(k_{y}b)\sin(l_{y}b)+\frac{P_{3}}{N}\cos(k_{z}c)\cos(l_{z}c)
+\frac{P_{4}}{N}\sin(k_{z}c)\sin(l_{z}c),\label{kernel}
\end{equation}
with coefficients given by
\begin{mathletters}
\label{allPs}
\begin{eqnarray}
P_{0}=&&U-V_{y}-V_{z}-J_{y}-J_{z}+(V_{x}+J_{x})(\cos(2k_{F}a)+1)
+2F_{x}+8C_{x}\cos(k_{F}a),\label{Pa}\\
P_{1}=&&-2F_{y}+J_{y}+V_{y},\label{Pb}\\
P_{2}=&&2F_{y}+J_{y}+V_{y},\label{Pc}\\
P_{3}=&&-2F_{z}+J_{z}+V_{z},\label{Pd}\\
P_{4}=&&2F_{z}+J_{z}+V_{z}.\label{Pe}
\end{eqnarray}
\end{mathletters}
As is seen above in Eq.(\ref{kernel}), the kernel turns out to be
diagonal on the basis of the leading harmonics on the $(k_y,k_z)$ plane.
Consequently, the gap will be of the form
\begin{equation}
\Delta({\bf k})=\Delta_{0}+\Delta_{1}\cos(k_{y}b)+
\Delta_{2}\sin(k_{y}b)+\Delta_{3}\cos(k_{z}c)+\Delta_{4}\sin(k_{z}c).
\label{gap}
\end{equation}
For vanishing order parameter the five components in (\ref{gapeq}) decouple
completely, and the critical temperature for the development of each type
of gap can easily be evaluated. For the conventional SDW with constant gap
we recover the well known result
\begin{equation}
k_{B}T_{c}^{(0)}=\frac{2\gamma}{\pi}v_{F}k_{F}e^{-2/P_{0}
\rho_0(0)},\label{Tc0}
\end{equation}
with $\gamma=1.781$, the Euler constant, and $\rho_0(0)=
a/\pi v_F$ is the electron density of states in the normal state
per spin at the Fermi surface. The critical temperature for the four kinds of
unconventional gap formation ($j=1,...,4$) is
\begin{equation}
k_{B}T_{c}^{(j)}=\frac{2\gamma}{\pi}v_{F}k_{F}e^{-4/P_{j}
\rho_0(0)}.\label{Tcj}
\end{equation}
On cooling the system, that type of an SDW
will develop first, for which the $T_c$ is the highest. The condition
for the formation of an USDW of type $j$ is $P_j>2P_0$. For example in case
of a half filled band an unconventional phase with the gap function
$\Delta({\bf k})=\Delta_2\sin(bk_y)$ will form if ${3\over 2}(V_y+J_y)+F_y
+V_z+J_z>U+2F_x$. Clearly, a combination of interchain Coulomb and exchange
integrals overwhelming the on site repulsion will facilitate the development
of an USDW (negative interchain pair-hopping integral favours a gap with
cosine dependence).

\begin{figure}%[t]
%\hskip 3truecm
\psfrag{kx}[t][b][1][0]{$a(k_x-k_F)$}
\psfrag{ky}[t][b][1][0]{$bk_y$}
\psfrag{E}[t][b][1][0]{$E_+\over\Delta_1$}
\includegraphics[width=8.7cm,height=7cm]{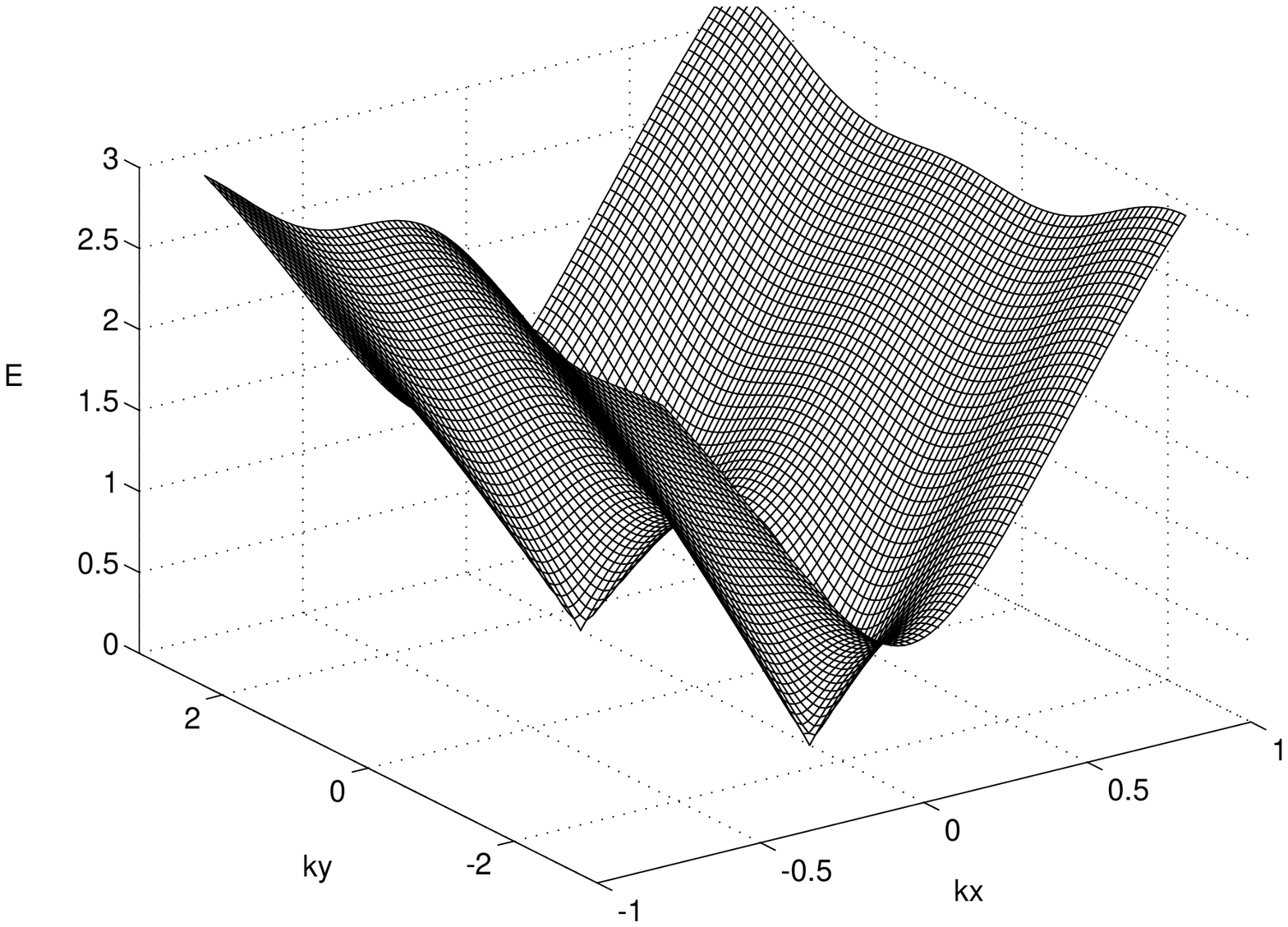}
%\psfrag{kx}[t][b][1][0]{$a(k_x-k_F)$}
%\psfrag{ky}[t][b][1][0]{$bk_y$}
\psfrag{E}[t][b][1][0]{$E_+\over\Delta_2$}
\includegraphics[width=8.7cm,height=7cm]{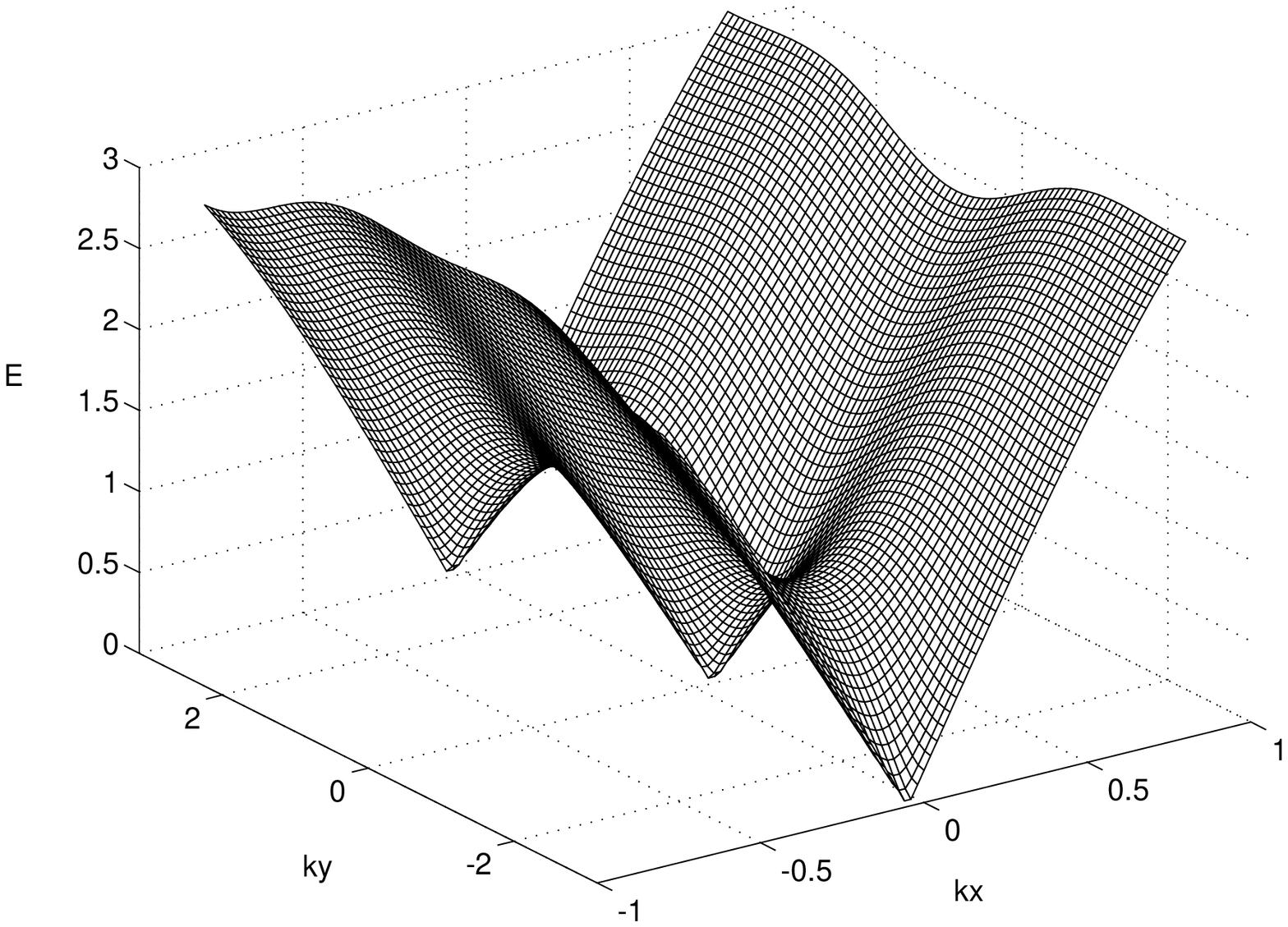}
%\vskip 1truecm
\caption{
Quasiparticle energy spectrum of an USDW. $E_+({\bf k})$ is given for
$\Delta({\bf k})=\Delta_1\cos(bk_y)$ (left panel), and for
$\Delta({\bf k})=\Delta_2\sin(bk_y)$ (right panel).
Other parameters are chosen as
$t_a/\Delta_{1,2}=2$, $t_b/\Delta_{1,2}=0.1$ and $ak_F=\pi/2$.
\label{fig:energy}}
\end{figure}

\begin{figure}%[h!]
\hskip 2truecm
\psfrag{x}[t][b][1][0]{$E/\Delta_{0,j}$}
\psfrag{y}[b][t][1][0]{$\rho(E)/\rho_0(0)$}
\psfrag{felso}{--- $\Delta_0$}
\psfrag{also}{\textperiodcentered\textperiodcentered\textperiodcentered
$\Delta_{j}$}
\includegraphics[width=13cm,height=7cm]{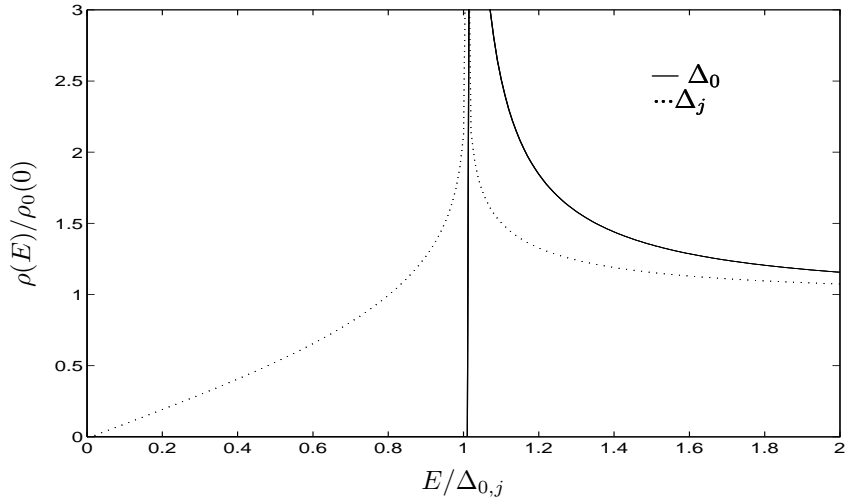}
\caption{The density of states for conventional (full line), and
unconventional (dotted line) density waves.
\label{fig:dos}}
\end{figure}

The quasiparticle energy spectrum from Eq.(\ref{Espectrum})
is shown in Fig.~\ref{fig:energy} for the two
typical unconventional gap functions of $k_y$ (we neglected any dispersion
in $k_z$ here for clarity). The excitation energy vanishes on lines (note the
additional $k_z$ direction in the Brillouin zone) of the Fermi surface,
and this will determine the nature of the thermodynamics of the system.
The corresponding density of states (DOS) is calculated as
\begin{equation}
{\rho(E)\over\rho_0(0)}=\int_{-\pi}^{\pi}{d(bk_y)\over 2\pi}
\int_{-\pi}^{\pi}{d(ck_z)\over 2\pi}{\rm Re}{|E|\over\sqrt{E^2-
|\Delta({\bf k})|^2}},\label{dos}
\end{equation}
and is shown in Fig.~\ref{fig:dos} for both the well known conventional
situation and for the unconventional cases determined analytically by
$\rho(E)/\rho_0(0)=(2|E|/\pi|\Delta_j|)K(|E|/|\Delta_j|)$ if $|E|<|\Delta_j|$,
and $\rho(E)/\rho_0(0)=(2/\pi)K(|\Delta_j|/|E|)$ if $|E|>|\Delta_j|$.
In the latter case the DOS
vanishes linearly at the Fermi energy and diverges logarithmically at the
maximum gap value, as follows from the properties of the complete elliptic
function of the first kind.

\begin{figure}%[h!]
\hskip 2truecm
\psfrag{t}[t][b][1][0]{$T/T_c$}
\psfrag{d}[b][t][1][0]{$\Delta(T)/\Delta(0)$}
\psfrag{felso}{--- $\Delta_0$}
\psfrag{also}{\textperiodcentered\textperiodcentered\textperiodcentered
$\Delta_{j}$}
\includegraphics[width=13cm,height=7cm]{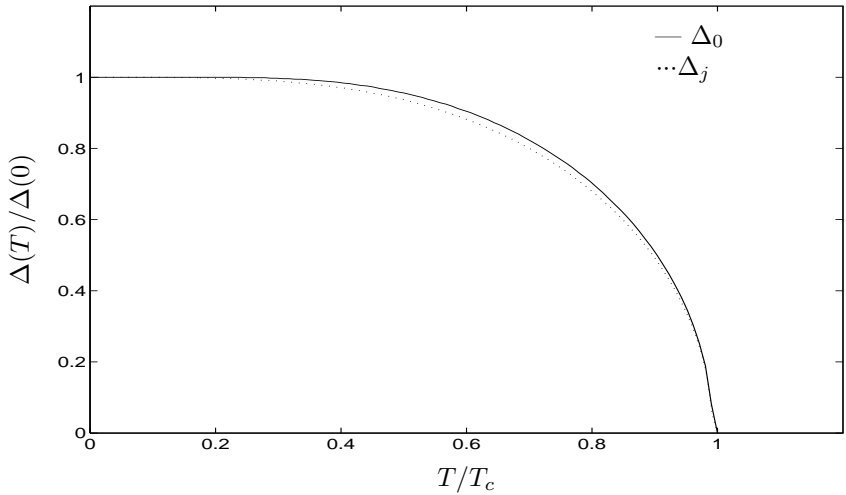}
\caption{The temperature dependence of the gap amplitude for conventional
($\Delta_0$) and unconventional ($\Delta_{j}$) density wave.
\label{fig:delta}}
\end{figure}

Assuming that only one kind of gap (either the conventional or one of the
four unconventionals, whichever opens at the highest $T_c$) persists all the
way down to zero temperature, we can use Eq.(\ref{gapeq}) in order to evaluate
the temperature dependence of the gap amplitude. This is the same for all
unconventional gap types, and is shown on Fig.~\ref{fig:delta} along with
the conventional dependence, displaying only small differences between the two.
At zero temperature the unconventional gap takes the value
$|\Delta_j(0)|=(2\pi/
\gamma\sqrt{e})k_BT_c^{(j)}$, leading to a gap maximum
to $T_c$ ratio of 4.28, instead of the ratio 3.52 in the conventional case.
For $T\ll T_c$ the unconventional gap decreases from its $T=0$ value
as $|\Delta_{j}(T)/\Delta_{j}(0)|=1-3\zeta(3)[k_BT/
|\Delta_{j}(0)|]^3$, which is
to be contrasted with the exponential correction for a constant gap. Similar
power law behavior is found for other quantities as well, due to the line
nodes in the excitation spectrum. For example the specific heat vanishes like
$T^2$ at low temperatures, and normalized to the normal state value we obtain
$C_v^{(j)}(T)/C_v^{(n)}(T)=[27\zeta(3)/\pi^2][k_BT/|\Delta_j(0)|]$,
as opposed to the
exponential freezing out for a conventional gap. Close to the transition
temperature the unconventional gap vanishes like
$[|\Delta_{j}(T)|/k_BT_c^{(j)}]^2=[32\pi^2/21\zeta(3)][1-T/T_c^{(j)}]$,
i.e. in
a square root manner with somewhat different prefactor compared to the
conventional case ($8\pi^2/7\zeta(3)$).
The mean field transition leads to a specific heat jump
at $T_c$ with the relative value of $\Delta C_v^{(j)}/C_v^{(n)}=8/7\zeta(3)
=0.95$, exactly two third of the conventional value.
Due to the presence of line nodes in the gap function the above thermodynamic
properties are identical to those of a $d_{x^2-y^2}$ superconductor\cite{Maki}
in spite of the different topology of their Fermi surfaces.

\section{Correlation functions and the nature of the order parameter}

At this point it is important to call the attention to the fact that while in
a conventional SDW the order parameter is proportional to the Fourier
component of the magnetization density at the nesting vector ${\bf Q}$, an
unconventional order parameter has nothing to do with the magnetization. In
order to see this we evaluate the magnetization
using the transformation in Eq.(\ref{d-a}):
\begin{equation}
<m({\bf Q})>=-\mu_{B}\sum_{\bf k,\sigma}^\prime
\sigma<a_{\bf k-Q,\sigma}^{+}a_{\bf
k,\sigma}>=\mu_{B}\sum_{\bf k}^\prime
\frac{\Delta({\bf k})\tanh\{{\beta\over 2}
E({\bf k})\}
 }{E({\bf k})}.\label{magnet}
\end{equation}
It is easily seen that Eq.(\ref{magnet}) yields zero magnetization for any of
the four unconventional gap functions, as opposed to the conventional
situation leading to $<m({\bf Q})>=2N\mu_{B}\Delta_0/P_0$. This means that an
USDW is {\it not} accompanied by a spatially periodic modulation of the spin
density, although the expectation value $<a_{\bf k-Q,\sigma}^{+}a_{\bf
k,\sigma}>$ becomes finite and
the existence of a robust thermodynamic phase transition
is unquestionable. This feature of the unconventional density waves makes them
suitable candidates for explaining low temperature phase transitions, where
conventional order parameters such as charge-, or spin-density modulation
are not observable, like in
$\alpha$-(BEDT-TTF)$_2$KHg(SCN)$_4$\cite{Christ}, or in
URu$_2$Si$_2$\cite{Isaacs} respectively.

What kind of physical quantity does then an unconventional gap correspond to?
Again utilizing the gap equation (\ref{gapeq}), we can convince
ourselves that if for example an unconventional gap
$\Delta_{1}\cos(k_{y}b)$ develops at low temperature, then the quantity
\begin{equation}
\tilde S_z({\bf q})={1\over 2}\sum_{\bf k,\sigma}
\sigma\sin[b(k_y+q_y/2)]a_{\bf k,\sigma}^{+}a_{{\bf k+q},\sigma}
\label{effspin}
\end{equation}
assumes a finite expectation value $<\tilde S_z({\bf Q})>=N\Delta_1/P_1$.
Therefore the Fourier component with wavenumber ${\bf Q}$ of the
"effective" spin density $\tilde S_z({\bf r})$ plays the role of the order
parameter in this unconventional case. Clearly, experimental observation of
this order parameter is possible only in probes coupling directly to this
physical quantity. The situation is rather similar to electronic Raman
scattering, where the photon-electron vertex carries momentum dependence,
and measures the effective density correlation function\cite{AbrGen,Multiband},
instead of scattering on just density fluctuations.

The rest of this chapter will be devoted to the evaluation of certain
correlation functions which are of particular interest. We begin with the
charge susceptibility
$\chi_{nn}({\bf q},t)=i<[n({\bf q},t),n({\bf -q},0)]>$, the autocorrelation
function of the density operator
$n({\bf q})=\sum_{\bf k,\sigma}a_{\bf k,\sigma}^{+}a_{\bf k+q,\sigma}$. The
quasiparticle contribution to the
frequency dependent charge susceptibility in the long wavelength
limit reads as
\begin{eqnarray}
\chi_{nn}({\bf q},\omega)&&=\frac{1}{V}\sum_{\bf k,\sigma}^\prime
\Bigg[\frac{1}{2}\Bigg(1+\frac
{\xi({\bf k})\xi({\bf k+q})+{\rm Re}(\overline{\Delta({\bf k})}
\Delta({\bf k+q}))}{E({\bf k})E({\bf k+q})}\Bigg)
\{f[E({\bf k+q})]-f[E({\bf k})]\}\times \nonumber \\
&&\times \Bigg(\frac{1}{\omega+
  E({\bf k})-E({\bf k+q})}-\frac{1}{\omega-
  E({\bf k})+E({\bf k+q})}\Bigg)+
\frac{1}{2}\Bigg(1-\frac
 {\xi({\bf k})\xi({\bf k+q})+{\rm Re}(\overline{\Delta({\bf k})}
 \Delta({\bf k+q}))}{E({\bf k})E({\bf k+q})}\Bigg)\times\nonumber\\
&&\times\{1-f[E({\bf k+q})]-f[E({\bf k})]\}
\Bigg(\frac{1}{\omega+
  E({\bf k})+E({\bf k+q})}-\frac{1}{\omega-
  E({\bf k})-E({\bf k+q})}\Bigg)\Bigg],\label{chi0}
\end{eqnarray}
while for wavenumbers around the nesting vector we obtain
\begin{eqnarray}
\chi_{nn}({\bf Q+q},\omega)=\frac{1}{4V}\sum_{\bf k,\sigma}^\prime
\Bigg[
 \frac{f[E({\bf k+q})]-f[E({\bf k})]}{\omega+
  E({\bf k})-E({\bf k+q})}
\Bigg(1-
 \frac{\xi({\bf k})}{E({\bf k})}\Bigg)
\Bigg(1+\frac{\xi({\bf k+q})}
 {E({\bf k+q})}
\Bigg)+
 \nonumber \\
 +
 \frac{f[E({\bf k})]-f[E({\bf k+q})]}{\omega-
  E({\bf k})+E({\bf k+q})}
\Bigg(1+
 \frac{\xi({\bf k})}{E({\bf k})}\Bigg)
\Bigg(1-\frac{\xi({\bf k+q})}
 {E({\bf k+q})}
\Bigg)+
\nonumber \\
 +
 \frac{1-f[E({\bf k})]-f[E({\bf k+q})]}{\omega+
  E({\bf k})+E({\bf k+q})}
\Bigg(1-
 \frac{\xi({\bf k})}{E({\bf k})}\Bigg)
\Bigg(1- \frac{\xi({\bf k+q})}
 {E({\bf k+q})}
\Bigg)+
\nonumber \\
+
 \frac{f[E({\bf k})]+f[E({\bf k+q})]-1}{\omega-
  E({\bf k})-E({\bf k+q})}
\Bigg(1+
 \frac{\xi({\bf k})}{E({\bf k})}\Bigg)
\Bigg(1+\frac{\xi({\bf k+q})}
 {E({\bf k+q})}
\Bigg)\Bigg].\label{chiQ}
\end{eqnarray}
The analytic structure of the spin susceptibilities are very similar to
Eqs.(\ref{chi0}) and (\ref{chiQ}). For example the longitudinal spin
susceptibility $\chi_{S_zS_z}=\chi_{nn}/4$
for all wavelengths, and the transverse spin
susceptibility $\chi_{S^+S^-}=\chi_{nn}/2$ for short wavelengths.
For long wavelengths however, we encounter
coherence factors different from those in Eq.(\ref{chi0}), namely the sign
of the two ${\rm Re}(\overline{\Delta({\bf k})}\Delta({\bf k+q}))$ terms
becomes negative.

In the followings we evaluate the above mentioned correlation functions
in both the static (first $\omega\rightarrow 0$, then ${\bf q}\rightarrow 0$),
and the dynamic (first ${\bf q}\rightarrow 0$, then $\omega\rightarrow 0$)
limit.

\subsection{Susceptibilities in the long wavelength limit}

\begin{figure}%[h!]
\hskip 2truecm
\psfrag{T}[t][b][1][0]{$T/T_c$}
\psfrag{CHI}[b][t][1][0]{$\chi_{nn}^{0S}(T)/g_0(0)$}
\psfrag{felso}{- - $\Delta_{j}$}
\psfrag{also}{\textperiodcentered\textperiodcentered\textperiodcentered 
$\Delta_0$}
\includegraphics[width=13cm,height=7cm]{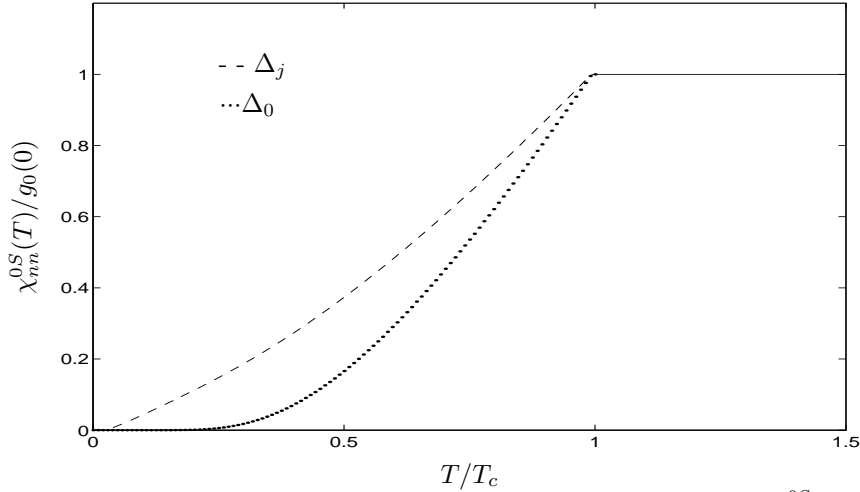}
\caption{The temperature dependence of the long wavelength static charge
susceptibility $\chi_{nn}^{0S}$ for conventional (dotted line) and for
unconventional (dashed line) density waves.
\label{fig:chin0s}}
\end{figure}

\begin{figure}%[h!]
\hskip 2truecm
\psfrag{T}[t][b][1][0]{$T/T_c$}
\psfrag{CHI}[b][t][1][0]{$\chi_{S^+S^-}^{0D}(T)/[g_0(0)/2]$}
\psfrag{felso}{- - $\Delta_{j}$}
\psfrag{also}{\textperiodcentered\textperiodcentered\textperiodcentered
$\Delta_0$}
\includegraphics[width=13cm,height=7cm]{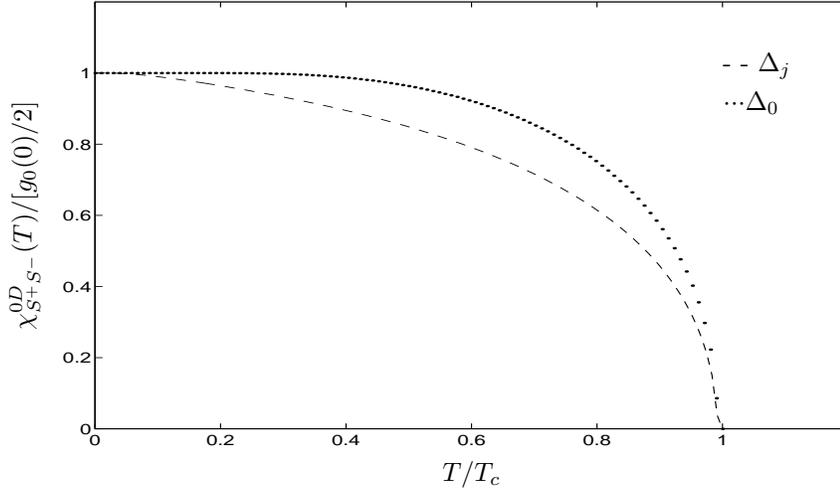}
\caption{The temperature dependence of the long wavelength dynamic
transverse spin
susceptibility $\chi_{S^+S^-}^{0D}$ for conventional (dotted line) and for
unconventional (dashed line) density waves.
\label{fig:chis0d}}
\end{figure}

The dynamic limit of the long wavelength charge susceptibility is
trivially zero, therefore we begin with the static limit of Eq.(\ref{chi0}):
\begin{equation}
\chi_{nn}^{0S}={2\over V}\sum_{{\bf k},\sigma}^\prime
\{-f^\prime[E({\bf k})]\}=g_0(0)\int_{-\infty}^\infty dE
[-f^\prime(E)]\rho(E)/\rho_0(0),\label{n0s}
\end{equation}
where $f^\prime(E)$ is the derivative of the Fermi function, and
$g_0(0)=2N\rho_0(0)/V$ is the normal state DOS (per unit volume)
at the Fermi energy
including spin degeneracy. The above susceptibility takes this value for
$T>T_c$, and its temperature dependence is shown in Fig.~\ref{fig:chin0s}
for both the conventional and the unconventional cases. We note here, that
following superconductivity terminology,
Eq.(\ref{n0s}) is often expressed as
$\chi_{nn}^{0S}=g_0(0)(1-f_s)$, where $f_s$ is the
superfluid fraction. In our case of course, condensate
fraction is
more appropriate. Due to the above mentioned relations between correlation
functions, Fig.~\ref{fig:chin0s} represents also the longitudinal spin
susceptibility, experimentally accessible through Knight-shift measurement,
vanishing at low temperatures. This is in contrast to the static homogeneous
transverse spin susceptibility $\chi_{S^+S^-}^{0S}=V^{-1}
\sum_{\bf k}^\prime|\Delta({\bf k})|^2/[E({\bf k})]^3=g_0(0)/2$, which
is independent of temperature. Returning to the charge (or longitudinal spin)
susceptibility, Eq.(\ref{n0s}) is evaluated at low temperatures yielding
exponential freezeout for conventional SDW, and
$\chi_{nn}^{0S}/g_0(0)=2\ln(2)k_BT/|\Delta_j(0)|$ for the USDW. Close to $T_c$
on the other hand $\chi_{nn}^{0S}/g_0(0)=(4T/T_c^{(j)}-1)/3$ for USDW (instead
of the conventional value $2T/T_c^{(0)}-1$).

Wrapping up our discussion of the long wavelength correlation functions, we
consider the dynamic limit of the transverse spin susceptibility
\begin{equation}
\chi_{S^+S^-}^{0D}={1\over V}
\sum_{\bf k}^\prime{|\Delta({\bf k})|^2\over [E({\bf k})]^3}
\{1-2f[E({\bf k})]\}={g_0(0)\over 2}\int_0^\infty dE{1-2f(E)\over E^2}
\int_{-\pi}^{\pi}{d(bk_y)\over 2\pi}\int_{-\pi}^{\pi}{d(ck_z)\over 2\pi}
{\rm Re}{|\Delta({\bf k})|^2\over\sqrt{E^2-|\Delta({\bf k})|^2}},\label{s0d}
\end{equation}
which is nonzero in the DW state, as shown in Fig.~\ref{fig:chis0d} for both
conventional and unconventional situations. The susceptibility normalized
to its zero temperature value
$\chi_{S^+S^-}^{0D}(T)/[g_0(0)/2]=f_d$
is again the condensate fraction, but now in the dynamic limit.
In order to
see the difference compared to the static condensate fraction,
we first realize that $[g_0(0)/4](f_d-f_s)=V^{-1}
\sum_{\bf k}^\prime\{-f^\prime[E({\bf k})]\}|\Delta({\bf k})|^2/
[E({\bf k})]^2$, then consider the
limiting values for low and high temperatures. For $T\rightarrow 0$ the
dynamic condensate fraction $f_d=1-\ln(2)k_BT/
|\Delta_j(0)|$ in USDW (the finite temperature correction is exponentially
small for the conventional case). Close to $T_c$ the dynamic condensate
fraction vanishes like $\pi|\Delta_0|/4k_BT$ for constant gap, and like
$|\Delta_j|/2k_BT$ for momentum dependent gap.

\subsection{Susceptibilities around the nesting vector}

We turn our attention now to the behavior of the short wavelength
correlation functions based on Eq.(\ref{chiQ}). These are of particular
interest in describing phenomena related to the phase transition involving
the nesting vector ${\bf Q}$. It is easily seen, that in the normal state
($T>T_c$) the charge susceptibility at the nesting vector follows a
logarithmic temperature dependence
\begin{equation}
\chi_{nn}^Q={1\over V}\sum_{{\bf k},\sigma}^\prime{1-2f[\xi({\bf k})]\over
2\xi({\bf k})}=
{g_0(0)\over 2}\ln{2\gamma v_Fk_F\over\pi k_BT}\label{cQn}
\end{equation}
for both conventional and unconventional cases independent of the
limiting procedure (static or dynamic).

\begin{figure}%[h!]
\hskip 2truecm
\psfrag{T}[t][b][1][0]{$T/T_c$}
\psfrag{CHI}[b][t][1][0]{$\chi_{nn}^{QS}(T)/\chi_{nn}^{QS}(T_c)$}
\psfrag{felso}{- - $\Delta_{j}$}
\psfrag{also}{\textperiodcentered\textperiodcentered\textperiodcentered
$\Delta_0$}
\includegraphics[width=13cm,height=7cm]{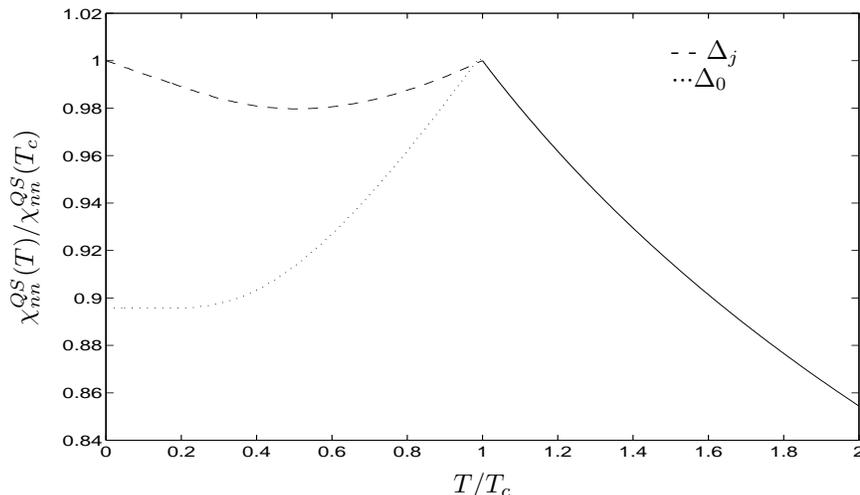}
\caption{The temperature dependence of the short wavelength charge
susceptibility $\chi_{nn}^Q$ in the
static limit for conventional (dotted line) and for
unconventional (dashed line) density waves.
We have chosen $P_0\rho_0(0)=P_j\rho_0(0)/2\approx 0.43$.
\label{fig:chinQs}}
\end{figure}

\begin{figure}%[h!]
\hskip 2truecm
\psfrag{T}[t][b][1][0]{$T/T_c$}
\psfrag{CHI}[b][t][1][0]{$\chi_{nn}^{QD}(T)/\chi_{nn}^{QD}(T_c)$}
\psfrag{felso}{- - $\Delta_{j}$}
\psfrag{also}{\textperiodcentered\textperiodcentered\textperiodcentered
$\Delta_0$}
\includegraphics[width=13cm,height=7cm]{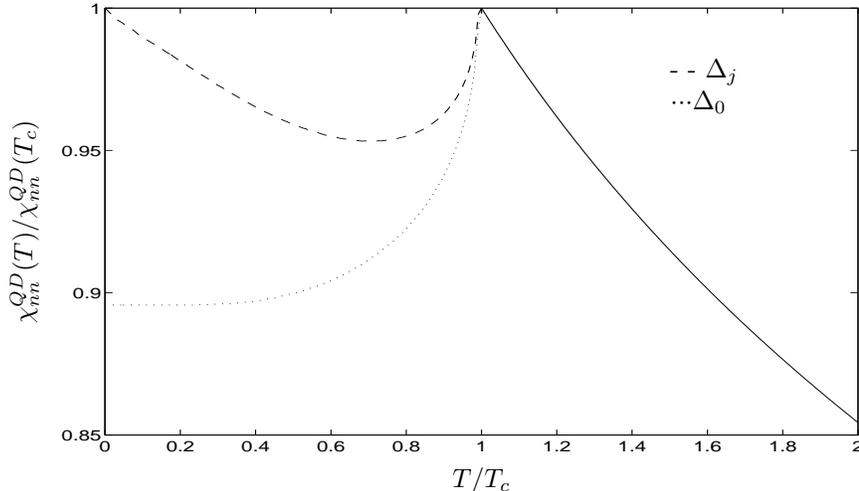}
\caption{The temperature dependence of the short wavelength charge
susceptibility $\chi_{nn}^Q$ in the
dynamic limit for conventional (dotted line) and for
unconventional (dashed line) density waves. Coupling constants are chosen
as in Fig.~\ref{fig:chinQs}.
\label{fig:chinQd}}
\end{figure}

Below the transition temperature we obtain a still rather simple formula
\begin{equation}
\chi_{nn}^Q={1\over V}\sum_{{\bf k},\sigma}^\prime{1-2f[E({\bf k})]\over
2E({\bf k})}-{g_0(0)\over 4}f,\label{cQdw}
\end{equation}
incorporating all the variations in the condensate fraction $f$, discussed
extensively in the previous subsection in all limits and cases. It is easy
to deal with the first term in Eq.(\ref{cQdw}) for a constant gap, since the
gap equation (\ref{gapeq}) relates just such an expression with the inverse
of the coupling constant. Therefore in the conventional case
$\chi_{nn}^Q=2N/VP_0-[g_0(0)/4]f$, leading to a monotonically decreasing
susceptibility in both the static and dynamic limit as the temperature is
lowered (see Figs.~\ref{fig:chinQs}~and~\ref{fig:chinQd}).

In case of a momentum dependent gap, the first term in Eq.(\ref{cQdw}) is
not exactly what appears in the gap equation (\ref{gapeq}), but at the expense
of a correction factor we can bring in the inverse coupling constant as in
the conventional case: $\chi_{nn}^Q=4N/VP_j-[g_0(0)/4](f-\delta f)$, where
\begin{equation}
\delta f=4\int_{-\pi}^{\pi}{d(bk_y)\over 2\pi}
\int_{-\pi}^{\pi}{d(ck_z)\over 2\pi}\int_0^{v_Fk_F}dE{\rm Re}{1-2f(E)\over
\sqrt{E^2-|\Delta({\bf k})|^2}}\left [{1\over 2}-
\left |{\Delta({\bf k})\over\Delta_j}\right |^2
\right ].\label{deltaf}
\end{equation}
This correction factor is evaluated as $\delta f=1-
4\ln(2)k_BT/|\Delta_j(0)|$ for low temperatures, and as $\delta f=(2/3)
(1-T/T_c^{(j)})$ close to the critical temperature.

Our results concerning the susceptibility at the nesting vector are
summarized in Figs.~\ref{fig:chinQs}~and~\ref{fig:chinQd} for the static and
dynamic limits respectively. In the conventional situation $\chi_{nn}^Q
\propto\chi_{S_zS_z}^Q$ is peaked at the critical temperature signaling the
phase transition. Indeed, if we consider the full spin susceptibility in
random phase approximation (RPA), the Stoner denominator vanishes exactly
if we approach $T_c^{(0)}$ from above,
leading to divergent response. The other (unconventional) coupling
channels do not contribute to the charge (spin) response due to their
momentum dependence. In the unconventional case however, RPA corrections
will not lead to divergence, since the dominant unconventional channel does not
couple to charge or spin density, while the conventional coupling constant is
not strong enough to make the Stoner denominator vanish. Instead, the
autocorrelation function of the effective spin density $\tilde S_z$ (see eg.
Eq.(\ref{effspin})) will be divergent at $T_c^{(j)}$ in RPA, as it should
if we are to have an unconventional phase transition.

\section{Optical conductivity}

The frequency dependent conductivity provides a wealth of information
about both the quasiparticle and the collective excitation spectrum of
density wave materials\cite{Gruner}. Therefore in this section we investigate
the properties of the conductivity tensor $\sigma_{\alpha\beta}(\omega)=
K_{\alpha\beta}({\bf q}=0,\omega)/i\omega$ in our model of USDW. The
electromagnetic kernel consists of so called paramagnetic and
diamagnetic parts: $K_{\alpha\beta}=K_{\alpha\beta}^P+K_{\alpha\beta}^D$.
The diamagnetic term
$K_{\alpha\beta}^D=-e^2V^{-1}\sum_{\bf k,\sigma}^\prime
m_{\alpha\beta}^{-1}({\bf k})$ involves the effective mass tensor derived
from the electronic spectrum (\ref{disp}). The paramagnetic part
$K_{\alpha\beta}^P=\chi_{j_\alpha j_\beta}$ is the
correlation function of the corresponding components of the current
operator given by ${\bf j}({\bf q})=-e\sum_{\bf k,\sigma}{\bf v}({\bf k})
a_{\bf k,\sigma}^{+}a_{\bf k+q,\sigma}$ in the ${\bf q}\rightarrow 0$ limit,
where the electron velocity is again
obtained from Eq.(\ref{disp}). The current correlation function turns out to
be given by the same formula as in Eq.(\ref{chi0}), except for the
different coherence factors and the multiplicative term
$e^2v_\alpha({\bf k})v_\beta({\bf k})$ under the summation. Since our system
is ideal in a sense that it does not include any source of damping for
electrons (for example impurity scattering), the real part of the conductivity
consists of a sharp Drude peak and a regular contribution:
${\rm Re}\sigma_{\alpha\beta}(\omega)=\pi D_{\alpha\beta}\delta(\omega)+
{\rm Re}\sigma_{\alpha\beta}^{reg}(\omega)$, where the Drude weight is given
by
\begin{equation}
D_{\alpha\beta}={e^2\over V}\sum_{\bf k,\sigma}^\prime
\left \{m_{\alpha\beta}^{-1}({\bf k})-v_\alpha({\bf k})v_\beta({\bf k})
{|\Delta({\bf k})|^2\over [E({\bf k})]^3}\{1-2f[E({\bf k})]\}\right\}.
\label{drude}
\end{equation}
The Drude peak is the only component in the normal state, but its weight
decreases below $T_c$ and vanishes completely at zero temperature. For
example in the chain direction the Drude wight is related to the dynamic
condensate fraction by $D_{xx}=e^2g_0(0)v_F^2(1-f_d)$. The
missing oscillator strength is taken over by the regular component at finite
frequencies:
\begin{equation}
{\rm Re}\sigma_{\alpha\beta}^{reg}(\omega)=g_0(0){\pi e^2\over\omega^2}
\tanh\left ({|\omega|\over 4k_BT}\right )\int_{-\pi}^{\pi}{d(bk_y)\over 2\pi}
\int_{-\pi}^{\pi}{d(ck_z)\over 2\pi}{\rm Re}{v_\alpha({\bf k})v_\beta({\bf k})
|\Delta({\bf k})|^2\over\sqrt{(\omega/2)^2-|\Delta({\bf k})|^2}}.
\label{sreg}
\end{equation}

\begin{figure}%[h!]
\hskip 2truecm
\psfrag{x}[t][b][1][0]{$\omega/|\Delta_0|$}
\psfrag{y}[b][t][1][90]{${\rm Re}\sigma_{\alpha\alpha}^{conv}
(\omega)2|\Delta_0|/
\pi e^2g_0(0)v_\alpha^2$}
{\rotatebox{-90}{\includegraphics[width=8.5cm,height=8cm]{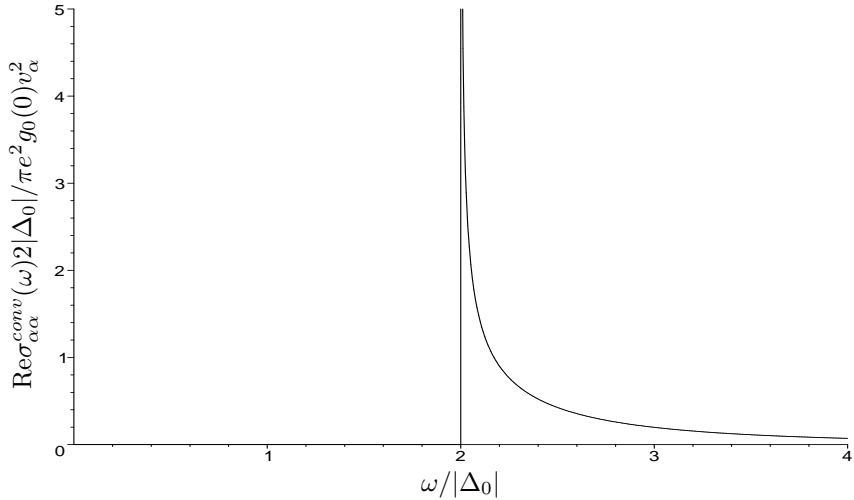}}}
\vskip -1.5truecm
\caption{The real part of the complex conductivity of a conventional SDW
at $T=0$.
\label{fig:sigconv}}
\end{figure}

In order to obtain characteristic lineshapes for the optical conductivity
tensor,
we first realize that it is diagonal, then consider various gap functions and
electric field directions. In case of a conventional SDW the conductivity is
given by
\begin{equation}
{\rm Re}\sigma_{\alpha\alpha}^{conv}(\omega>0)=e^2g_0(0)v_\alpha^2\tanh\left
({\omega\over 4k_BT}\right ){\pi|\Delta_0|^2\over\omega^2}{\rm Re}{1\over
\sqrt{(\omega/2)^2-|\Delta_0|^2}},\label{sigconv}
\end{equation}
where $v_x=v_F$, $v_y=\sqrt{2}bt_b$ and $v_z=\sqrt{2}ct_c$. The frequency
dependence of the conductivity is the same for all three directions of the
electric field, and is shown in Fig.~\ref{fig:sigconv} at zero temperature.
We recognize the sharp onset of absorbtion at $\omega=2|\Delta_0|$ due to the
constant gap.

\begin{figure}%[h!]
\hskip 2truecm
\psfrag{x}[t][b][1][0]{$\omega/|\Delta_j|$}
\psfrag{y}[b][t][1][90]{${\rm Re}\sigma_{xx}^{unc}(\omega)2|\Delta_j|/
e^2g_0(0)v_F^2$}
{\rotatebox{-90}{\includegraphics[width=8.5cm,height=8cm]{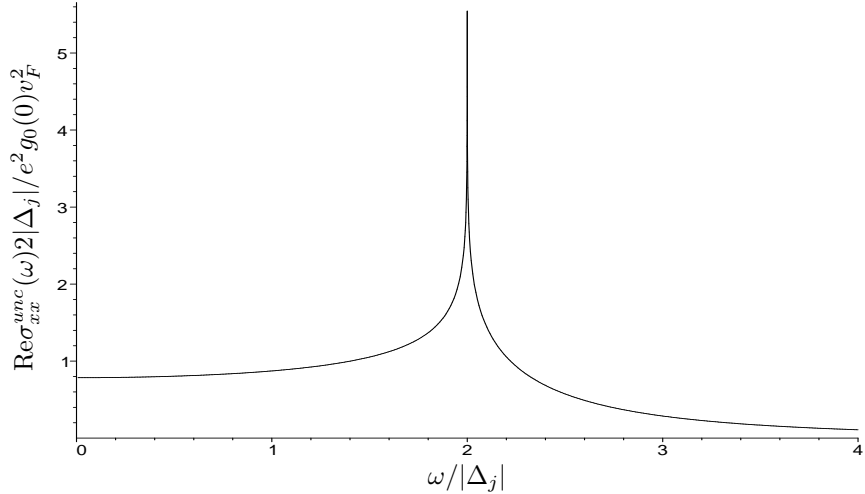}}}
\vskip -1.5truecm
\caption{The real part of the complex conductivity of an USDW in the chain
direction at $T=0$.
\label{fig:sigux}}
\end{figure}

In the unconventional situations we expect finite absorbtion below the
maximum gap, since we have low energy optical excitations around the nodes
of the order parameter. Consider first the conductivity in the chain direction
\begin{equation}
{\rm Re}\sigma_{xx}^{unc}(\omega>0)=e^2g_0(0)v_F^2\tanh\left
({\omega\over 4k_BT}\right ){4|\Delta_j|^2\over\omega^3}I\left ({2|\Delta_j|
\over\omega}\right ),\label{sigux}
\end{equation}
where $I(k)=\int_0^{\pi/2}d\varphi\sin^2\varphi{\rm Re}
(1-k^2\sin^2\varphi)^{-1/2}$, evaluated as
$I(k<1)=[K(k)-E(k)]/k^2$, and $I(k>1)=[K(1/k)-E(1/k)]/k$. This function
is plotted in Fig.~\ref{fig:sigux} at zero temperature. We see a logarithmic
divergence at the maximum optical gap, and a substantial absorbtion below that
gap as expected. We note here, that if we consider the conductivity in
one of the directions perpendicular to the chain, we obtain the same lineshape
if the unconventional gap varies in the other perpendicular direction. Of
course we need to replace $v_F$ in Eq.(\ref{sigux}) by the proper
perpendicular velocity. At this point it is appropriate to remark that we are
calculating only quasiparticle contributions to the conductivity, although
it is known that due to the sliding motion of the condensate a collective
contribution will also be observed in the chain direction\cite{VM}. However,
if the electric field is perpendicular to the chains, no such contribution is
expected, therefore our results are directly applicable to the experimental
situation.

\begin{figure}%[h!]
\hskip 2truecm
\psfrag{x}[t][b][1][0]{$\omega/|\Delta_2|$}
\psfrag{y}[b][t][1][90]{${\rm Re}\sigma_{yy}^{sin}(\omega)4|\Delta_2|/
e^2g_0(0)v_y^2$}
{\rotatebox{-90}{\includegraphics[width=8.5cm,height=8cm]{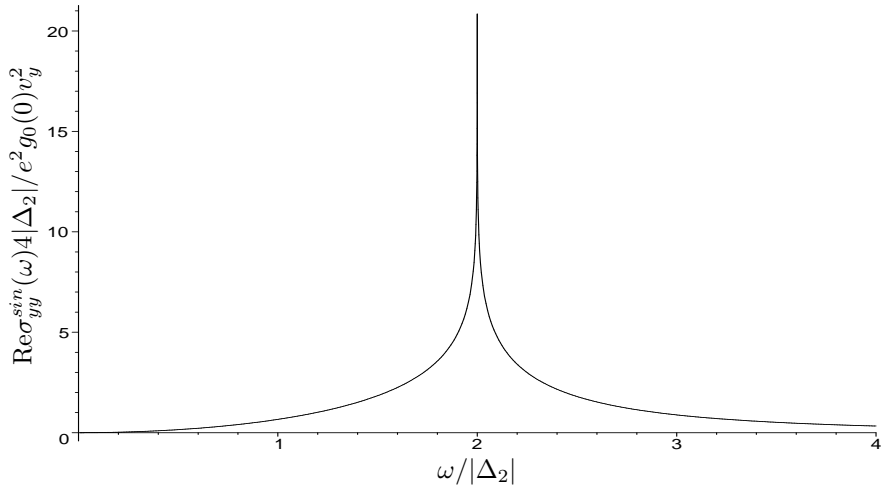}}}
\vskip -1.5truecm
\caption{
The real part of the complex conductivity of an USDW in
the $y$ direction at $T=0$, $\Delta({\bf k})=\Delta_2\sin(k_yb)$.
\label{fig:sigsin}}
\end{figure}

\begin{figure}%[h!]
\hskip 2truecm
\psfrag{x}[t][b][1][0]{$\omega/|\Delta_1|$}
\psfrag{y}[b][t][1][90]{${\rm Re}\sigma_{yy}^{cos}(\omega)4|\Delta_1|/
e^2g_0(0)v_y^2$}
{\rotatebox{-90}{\includegraphics[width=8.5cm,height=8cm]{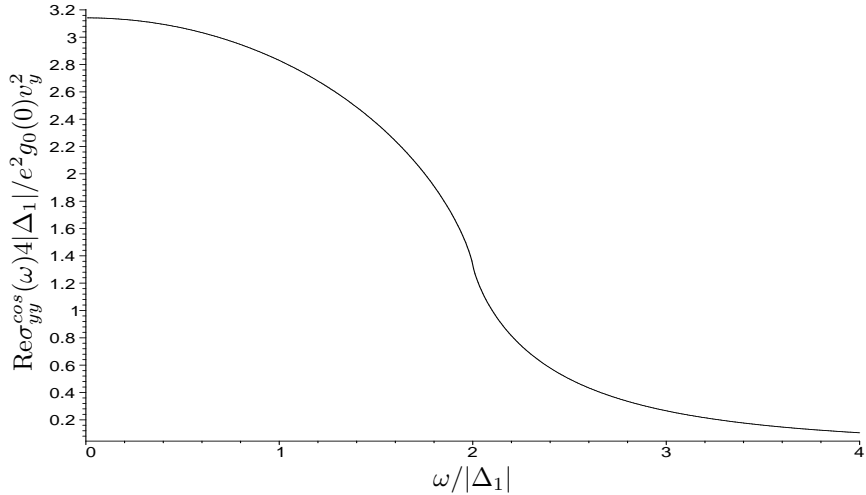}}}
\vskip -1.5truecm
\caption{
The real part of the complex conductivity of an USDW in
the $y$ direction at $T=0$, $\Delta({\bf k})=\Delta_1\cos(k_yb)$.
\label{fig:sigcos}}
\end{figure}

Finally, we consider the situation when the electric field is aligned in
that perpendicular direction in which the order parameter varies. Without
loss of generality we can take this to be the $y$ direction. There are two
possibilities corresponding to the gap function $\Delta({\bf k})=\Delta_1
\cos(k_yb)$, and $\Delta({\bf k})=\Delta_2\sin(k_yb)$. Eq.(\ref{sreg}) leads
to the following expressions for the conductivities:
\begin{equation}
{\rm Re}\sigma_{yy}^{cos}(\omega>0)=e^2g_0(0)v_y^2\tanh\left ({\omega\over
4k_BT}\right ){8|\Delta_1|^2\over\omega^3}I_{cos}\left ({2|\Delta_1|\over
\omega}\right ),\label{sigcos}
\end{equation}
and
\begin{equation}
{\rm Re}\sigma_{yy}^{sin}(\omega>0)=e^2g_0(0)v_y^2\tanh\left ({\omega\over 
4k_BT}\right ){8|\Delta_2|^2\over\omega^3}I_{sin}\left ({2|\Delta_2|\over
\omega}\right ),\label{sigsin}
\end{equation}
where $I_{sin}(k)=\int_0^{\pi/2}d\varphi\sin^4\varphi{\rm Re}
(1-k^2\sin^2\varphi)^{-1/2}$, and $I_{cos}(k)=I(k)-I_{sin}(k)$. The former
function is evaluated as $I_{sin}(k<1)=[(2+k^2)K(k)-2(1+k^2)E(k)]/3k^4$, and 
$I_{sin}(k>1)=[(1+2k^2)K(1/k)-2(1+k^2)E(1/k)]/3k^3$.
We plot these results in Fig.~\ref{fig:sigsin} and
in Fig.~\ref{fig:sigcos} for the sine and cosine dependence of the order
parameter respectively. As seen on Fig.~\ref{fig:sigsin},
due to the matching of the ${\bf k}$ dependences of the gap and the velocity
$v_y({\bf k})=2bt_b\sin(k_yb)$ the low frequency conductivity is 
suppressed and proportional to $\omega^2$, while the logarithmic
divergence is still there at $\omega=2|\Delta_2|$, as for the chain direction
conductivity (see Fig.~\ref{fig:sigux}). This
reasoning is quite similar to the one used in explaining the lineshape of the
$B_{1g}$ Raman response in $d_{x^2-y^2}$ superconductors \cite{Dev}.

In case of a cosine gap on the other hand, the velocity is zero at the gap
maximum, therefore the logarithmic divergence is cut off at
$\omega=2|\Delta_1|$, and we have a monotonically
decreasing conductivity as shown in Fig.~\ref{fig:sigcos}.

\section{Conclusions}

In this paper we have developed the mean field theory of unconventional
density waves
in quasi-one dimensional interacting electron systems. Although our
calculations refer explicitly to unconventional spin density waves (USDW),
most of
the results apply to unconventional charge density waves as well. We have
found that the excitation spectrum and thermodynamics of our model are
identical to those of a $d$-wave superconductor, due to the presence of line
nodes in the gap function $\Delta(k_y,k_z)$. Formation of an USDW is
facilitated by a combination of interchain Coulomb and exchange
integrals overwhelming the on site repulsion.

It is important to realize that an USDW is not accompanied by a
spatially periodic modulation of the spin density. Instead, an effective
spin density plays the role of the order parameter, which is observable only
in probes coupling to this quantity through form factors significantly
dependent on wavenumber. This feature of the unconventional density waves
makes them
suitable candidates for explaining low temperature phase transitions, where
conventional order parameters such as charge-, or spin-density modulation
are not observable. This state of affairs is sometimes referred to as 
"hidden order". The above conclusion is also corroborated by our investigation
of charge and spin correlation functions at the nesting vector, which do not
diverge at the critical temperature in random phase approximation,
unlike the effective spin density correlator, which does. The 
homogeneous static spin susceptibility
shows qualitatively the same anisotropic behavior below the transition
temperature as in the conventional case.

Finally, we have calculated the quasiparticle contribution to the
frequency dependent conductivity for an
USDW system without quasiparticle damping (collisionless limit). The
lineshape always exhibits absorbtion for frequencies below the maximum
optical gap in the quasiparticle spectrum, but
varies significantly depending on the direction of the electric
field and on the functional form of the gap function. These differences
can be exploited in determining the nature of the condensate by optical
spectroscopic tools.

\acknowledgments

This work was supported by the Hungarian National Research Fund under
grant numbers OTKA T032162 and T029877,
and by the Ministry of
Education under grant number FKFP 0029/1999.

%\bibliography{usdw}

%\end{multicols}    % for twocolumn
\end{document}